\documentclass[prl,showpacs,superscriptaddress,twocolumn]{revtex4}

\usepackage{amsmath,bm}
\usepackage{graphicx}
\usepackage{epsfig}

\begin{document}

\newcommand{\vk}{{\vec k}}
\newcommand{\vK}{{\vec K}}
\newcommand{\vb}{{\vec b}}
\newcommand{{\vp}}{{\vec p}}
\newcommand{{\vq}}{{\vec q}}
\newcommand{\vQ}{{\vec Q}}
\newcommand{\vx}{{\vec x}}
\newcommand{\beq}{\begin{equation}}
\newcommand{\eeq}{\end{equation}}
\newcommand{\half}{{\textstyle \frac{1}{2}}}
\newcommand{\gton}{\stackrel{>}{\sim}}
\newcommand{\lton}{\mathrel{\lower.9ex \hbox{$\stackrel{\displaystyle<}{\sim}$}}}
\newcommand{\ee}{\end{equation}}
\newcommand{\ben}{\begin{enumerate}}
\newcommand{\een}{\end{enumerate}}
\newcommand{\bit}{\begin{itemize}}
\newcommand{\eit}{\end{itemize}}
\newcommand{\bc}{\begin{center}}
\newcommand{\ec}{\end{center}}
\newcommand{\bea}{\begin{eqnarray}}
\newcommand{\eea}{\end{eqnarray}}

\newcommand{\beqar}{\begin{eqnarray}}
\newcommand{\eeqar}[1]{\label{#1} \end{eqnarray}}
\newcommand{\pleft}{\stackrel{\leftarrow}{\partial}}
\newcommand{\pright}{\stackrel{\rightarrow}{\partial}}

\newcommand{\eq}[1]{Eq.~(\ref{#1})}
\newcommand{\fig}[1]{Fig.~\ref{#1}}
\newcommand{\eff}{ef\!f}
\newcommand{\alphas}{\alpha_s}

\renewcommand{\topfraction}{0.85}
\renewcommand{\textfraction}{0.1}
\renewcommand{\floatpagefraction}{0.75}

\title{ Momentum Imbalance of Isolated  Photon-Tagged Jet Production at RHIC and LHC }

\date{\today  \hspace{1ex}}
\author{Wei Dai}

\affiliation{Key Laboratory of Quark \& Lepton Physics (MOE) and Institute of Particle Physics,
 Central China Normal University, Wuhan 430079, China}

\author{Ivan Vitev}
\affiliation{Los Alamos National Laboratory, Theoretical Division,
MS B238, Los Alamos, NM 87545, USA}

\author{Ben-Wei Zhang}
\email{bwzhang@iopp.ccnu.edu.cn}
\affiliation{Key Laboratory of Quark \& Lepton Physics (MOE) and Institute of Particle Physics,
 Central China Normal University, Wuhan 430079, China}

\begin{abstract}
In collisions of ultra-relativistic nuclei, photon-tagged jets provide
a unique opportunity to compare jet production and modification
due to parton shower formation and  propagation in strongly-interacting
matter at vastly different center-of-mass energies. We present first
results for the  cross sections of jets tagged by an isolated photon
to ${\cal O}(\alpha_{\rm em} \alpha_s^2)$ in central Au+Au reactions with
$\sqrt{s_{NN}}=200$~GeV at RHIC and central Pb+Pb reactions with
$\sqrt{s_{NN}}=2.76$~TeV  at LHC. We evaluate the increase in the
transverse momentum imbalance of the observed  $\gamma$+jet state, induced by
the dissipation of the parton shower energy due to strong final-state
interactions.
Theoretical predictions to help interpret recent and upcoming experimental
data are presented.
\end{abstract}


\maketitle

Hard parton scattering  processes of $Q^2 \gg \Lambda_{\rm QCD}^2$,
such as the ones that produce energetic and/or massive
 hadronic final states at the world's premier
collider facilities~\cite{Owens:1986mp},  have been widely used in the past
decade to investigate the  properties of the strongly-interacting
quark-gluon plasma (QGP) created  in ultra-relativistic  heavy-ion collisions.
Their utility as tomographic probes   of the QGP~\cite{Wang:1991xy} arises from the short
interaction time $\sim 1/Q$ and the clear separation between the
typical energy and momentum scales that characterize the
probe and the medium~\cite{Ovanesyan:2011xy}. Among such hard probes,
large transverse momentum jets~\cite{Sterman:1977wj,Ellis:1990ek}
have recently emerged as an new tool,
which not only provides the most promising channel to test and advance
perturbative Quantum Chromodynamics (QCD) in the many-body
environment of ultra-relativistic nuclear collisions,
but is also more sensitive to the details of the in-medium parton shower dynamics
than leading hadron production~\cite{Vitev:2008rz,Renk}.

Jets tagged by photons ($\gamma$) or electroweak bosons ($W^\pm, \, Z^0$)
are particularly well suited to studying heavy-ion
collisions~\cite{Neufeld:2010fj,Neufeld:2012df} since the tagging particle
escapes the region of strongly-interacting matter unscathed.
For example, the CMS collaboration measurements in lead-lead (Pb+Pb)
collisions show absence of significant modification  of
high transverse momentum  photon production relative to the
binary collision-scaled proton-proton (p+p) result within the current
statistical and systematic uncertainties~\cite{CMS:2012IP}.
Thus, in the collinear factorization approach $\gamma$s  can provide,
on average, constraints on the energy of the away-side parton shower~\cite{Wang:1997pj}.
Furthermore, jets tagged by photons or electroweak bosons are largely
unaffected by the fluctuations of the soft hadronic  background~\cite{Neufeld:2012df}
that may complicate the interpretation of di-jet modification in heavy-ion
collisions~\cite{Cacciari:2011tm,He:2011d,He:2012}. By selecting
a suitable range for the transverse momentum of the tagging
photon ($p_{T_\gamma}$), accessible at both RHIC and LHC, the in-medium
modification of parton showers in dense matter created at very different
$\sqrt{s_{NN}}$ can be studied. In the jet suppression region, where
the transverse momentum of the jet $p_{T_{\rm jet}} \geq p_{T_\gamma}$, the
attenuation of photon-tagged jets and inclusive jets~\cite{jets2010prl}
can be directly compared.

In addition to being theoretically attractive, the $\gamma$+jet channel
is being actively investigated by the LHC experiments. Recently, the ATLAS
collaboration reported progress in the measurement of photon+jet production
cross sections in p+p collisions at $\sqrt{s}=7$~TeV~\cite{ATLAS:2012ar}.
The CMS collaboration  observed a
significant enhancement in the transverse momentum imbalance of
photon-tagged jet events  central Pb+Pb reactions at $\sqrt{s_{NN}}=2.76$~TeV, providing a direct
quantitative measure of parton energy loss in the QCD
medium~\cite{Chatrchyan:2012vq}. Photon-tagged jet results are also expected in the
near future from RHIC experiments at much lower $\sqrt{s_{NN}}$.

With this motivation, we present theoretical predictions for the cross section
modification and the growth in the transverse momentum imbalance of photon-tagged
jet events in heavy ion collisions at RHIC and LHC. Our results combine
the ${\cal O}(\alpha_{\rm em} \alpha_s^2)$ perturbative cross sections with
initial-state cold nuclear matter effects and final-state parton shower modification
and energy dissipation in the QGP.


The cross section for prompt photon production is given by the sum of its
direct and fragmentation contributions and can be written
schematically as~\cite{Jetphox:2002, Jetphox:2009}:
\begin{eqnarray}
d\sigma^\gamma &=& d\sigma^{(D)}(\mu_R,\mu_f) \nonumber\\
&& +\sum_{k=q,\bar{q},g} d\sigma_k^{(F)}(\mu_R,\mu_f,\mu_{fr})
\otimes D_{\gamma/k}(\mu_{fr}) \, , \;
\label{sch}
\end{eqnarray}
Here, $\sigma^{(D)}$ (direct), $\sigma^{(F)}$ (fragmentation) are the corresponding
cross sections for a photon or a parton  and  $D_{\gamma/k}(\mu_{fr})$ is
the fragmentation function of parton $k$ into a photon.
In Eq.~(\ref{sch}) $\mu_R,\mu_f,\mu_{fr}$
are the renormalization, factorization and fragmentation scales, respectively.
We denote by $\otimes$ the standard convolution over the fragmentation
momentum fraction. In this paper we take advantage of JETPHOX~\cite{Jetphox:2002},
a Monte Carlo program  designed to calculate $p+p \rightarrow \gamma/h + J +X$.
For  inclusive $\gamma/h$ production it yields next-to-leading (NLO) order accuracy.
For the more differential $\gamma + {\rm jet}$ channel the   ${\cal O}(\alpha_{\rm em} \alpha_s^2)$
evaluation gives the leading cross section contribution away from
$p_{T_\gamma} \approx  p_{T_{\rm jet}}$.

TeV collider experiments, such as  CDF and D0 at the Tevatron, ATLAS and CMS at LHC,
implement constraints on the hadronic activity that accompanies photon candidate events.
By imposing an upper limit on the hadronic transverse energy in a given cone of radius
$ R_{\rm iso.} = \sqrt{(y-y_\gamma)^2+(\phi-\phi_\gamma)^2}$
around the photon direction, the cross section for isolated photons can be obtained.
The isolation cut not only rejects the background of secondary photons coming
from several decay channels, but also affects the prompt photon cross section itself
 by reducing the contributions of the fragmentation component~\cite{Jetphox:2009}.
Here, we will focus on jets tagged by isolated photons.

\begin{figure}[!t]
\begin{center}
\vspace*{-0.2in}
\hspace*{-.1in}
\includegraphics[width=3.4in,height=3.6in,angle=0]{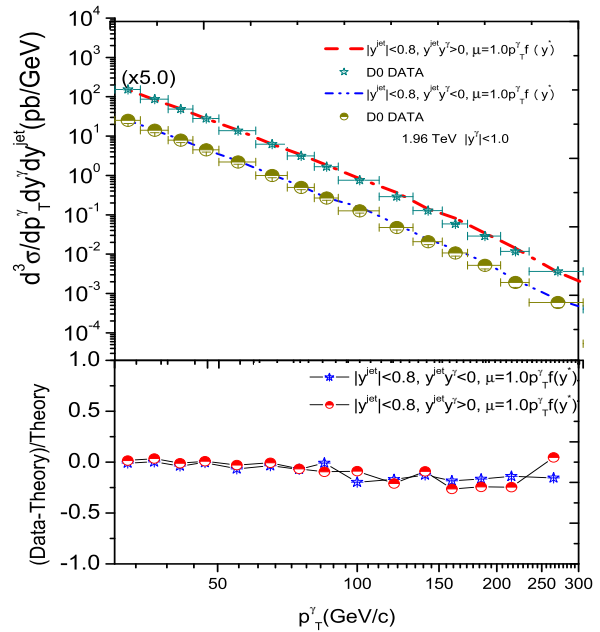}
\vspace*{-.2in}
\caption{Comparison between the isolated photon+jet cross section measured by
the D0 experiment and the  ${\cal O}(\alpha_{\rm em} \alpha_s^2) $ pQCD theory
in $\rm{p+\bar{p}}$ collisions at $\sqrt{s_{NN}}=1.96$~TeV.}
\label{fig:illustpp}
\end{center}
\end{figure}

Measurements of the production cross section of  isolated photons associated
with jets in proton-antiproton ($\rm{p+\bar{p}}$) collisions at  $\sqrt{s} = 1.96$~TeV
were performed by the D0 collaboration at the Tevatron~\cite{D0:2008}. The baseline
simulation in ``elementary'' nucleon-nucleon interactions can be validated against
the experimental results, as shown in
Fig.~\ref{fig:illustpp}.Here, the differential cross section of the final state
photon $d^3\sigma/dy_{\gamma} dy_{\rm jet} dp_{T_\gamma}$
is  given for specific transverse momentum  and rapidity cuts for the leading away-side jet.
We consider kinematic regions around mid rapidity $\vert y^{\gamma} \vert < 1.0$, $\vert y^{\text{jet}} \vert < 0.8$. 
The transverse momentum cuts are:
$p_{T_{\gamma}}>30$~GeV, $p_{T_{\rm jet}}>15$~GeV.
Our theoretical simulation employs
 the CTEQ 6.6 parton distribution functions and the BGF II
fragmentation functions~ \cite{Jetphox:2002}.
All scales are chosen to be
$\mu_{R,f,fr}=p_{T_\gamma}\sqrt{(1+exp(-2\vert y^* \vert)/2)}$, with $y^*=
0.5(y_\gamma-y_{\text{jet}})$. To compare to the D0 results,
we implement a midpoint cone algorithm
with a radius parameter $R=0.7$ for the jet and impose an isolation cut that
requires the hadronic transverse energy in the cone of radius $R_{\rm iso.}=0.4$ around
the photon direction  to be less than 7\%  of the $\gamma$ energy.
We find that  the theoretical predictions provided by JETPHOX  are in
good agreement with the measured cross section within the statistical and
systematic uncertainties for either equal ($y_\gamma y_{\rm jet}>0$)  or different
($y_\gamma y_{\rm jet} < 0$) signs of the jet and photon rapidities. The bottom panel of
Fig.~1 shows the fractional deviation (Data-Theory)/Theory.

In reactions with heavy nuclei, cold nuclear matter (CNM) effects prior to the QGP formation,
such as initial-state energy loss, power corrections and the Cronin effect, can modify the
experimentally observed photon+jet cross sections. At  RHIC and LHC  we are
interested in jets of $p_{T_{\rm jet}}>10$~ GeV or 30~GeV, respectively. At these energy scales
only  initial-state energy loss may play a role~\cite{Vitev:2007ve,He:2012}.
As we will see below, the CNM effects do not affect the asymmetry of
photon+jet events. The suppression of photon+jet production,
induced by CNM effect, is fairly small.
Final-state quark-gluon plasma effects include radiative energy loss,
caused by medium-induced parton splitting~\cite{Vitev:2007ve}, and the
dissipation of the energy of the parton
shower through collisional interactions in the strongly-interacting
matter~\cite{Neufeld:2011yh}. To evaluate the
isolated photon+jet cross section, we use the theoretical framework
of~\cite{Vitev:2008rz,jets2010prl}, which clearly outlines the  similarities and
differences between jet and leading particle production in the ambiance of the QGP.
The medium-modified photon+jet cross section per binary scattering is calculated as
follows:
\begin{eqnarray}
\frac{1}{\langle N_{bin}\rangle}\frac{d\sigma^{AA}}{dp_{T_\gamma}dp_{T_{\rm jet}}}
&=& \sum_{q,g} \int_0^1 d\epsilon \frac{P_{q,g}(\epsilon)}{1-[1-f(R)]\epsilon}
\nonumber\\
 && \hspace*{-2.5cm}  \times R_{q,g} \,
\frac{d\sigma^{CNM} \left(p_{T_\gamma},\frac{p_{T_{\rm jet}}}{1-[1-f(R)]\epsilon }\right)  }
{dp_{T_\gamma}dp_{T_{\rm jet}}} \, .
\label{eq:modify}
\end{eqnarray}
The $\gamma$ + jet production points are distributed according to the binary
nucleon-nucleon collision density and propagate through the medium that follows
the participant number density and undergoes Bjorken expansion. The  
properties of the gluon-dominated medium are related to the local temperature: 
$m_D = g_{\rm med}T$, $\sigma^{gg} = $,  $\lambda = 1/(\sigma^gg \rho)$ 
In Eq.~(\ref{eq:modify}) $P_{q,g}(\epsilon)$ is the probability distribution that a
fraction $\epsilon$ of the hard-scattered quark or gluon energy is converted to
a medium-induced parton shower~\cite{Vitev:2008rz, Neufeld:2010fj}  and $R_{q,g}$
is the fraction of the corresponding  hard-scattered partons.  Part of the
dependence of the jet cross section on the jet  reconstruction parameters,
such as the radius $R$, is contained  in $d\sigma^{CNM}/dp_{T_\gamma}dp_{T_{\rm jet}}$.
More importantly, the fraction of the parton
shower energy that is simply redistributed inside the jet due to final-state
interactions $f(R)$ also depends on $R$  [$f(R)_{R\rightarrow 0} \rightarrow 0$,
$f(R)_{R \gg 1} \rightarrow 1$] ~\cite{Vitev:2008rz, jets2010prl,He:2011d}.
In out calculation $P_{q,g}(\epsilon)$ and  $f(R)$ are evaluated on an
event-per-event basis.
The physics meaning of Eq.~(\ref{eq:modify}) is that the observed photon-tagged
jet cross section in nucleus-nucleus reactions is a probabilistic superposition of
cross sections where the jet is of higher initial transverse momentum
$ p_{T_{\rm jet}}/\{1-[1-f(R)]\epsilon) \}$.

\begin{figure}[!t]
\begin{center}
\vspace*{-0.15in}
\includegraphics[width=3.2in,height=2in,angle=0]{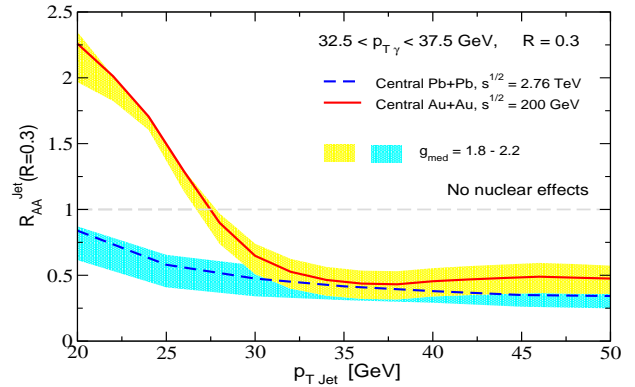}
\vspace*{-.1in}
\caption{The nuclear modification factor $R_{AA}^{\gamma-\rm jet}$ for the
isolated photon-tagged jets in the most central
A+A reactions at RHIC and LHC  (R=0.3).}
\label{fig:raa}
\end{center}
\end{figure}

The many-body QCD dynamics that modifies isolated photon + jet production in
relativistic heavy-ion collisions is manifested in the deviation from the
baseline p+p results, scaled by the number of binary nucleon-nucleon
interactions $\langle N_{bin}\rangle$~\cite{Vitev:2008rz,Neufeld:2012df}:
\begin{eqnarray}
 R_{AA}^{\gamma-{\rm jet}}(p_{T_{\rm jet}},p_{T_\gamma};R) =
\frac{\frac{d\sigma^{AA}}{dp_{T_\gamma}dp_{T_{\rm jet}}}}{\langle N_{bin}\rangle
\frac{d\sigma^{pp}}{dp_{T_\gamma}dp_{T_{\rm jet}}}} \; .
\end{eqnarray}
Theoretical predictions for the nuclear modification of the photon+jet production
rate are presented in Fig.~\ref{fig:raa}, where we show
$R_{AA}^{\rm jet}(p_{T_{\rm jet}};R)$ with the tagging $\gamma$ momentum integrated
in the region  $ 32.5  <  p_{T \gamma } < 37.5 $~GeV. We choose a jet reconstruction
radius  R=0.3  and  give  results for central Au+Au collisions at RHIC
(solid line) central Pb+Pb collisions at LHC (dashed line).
We note direct comparison between RHIC and LHC is possible for the first time in a
more exclusive channel. The largest suppression is  observed for
  $p_{T_{\rm jet}} \approx  p_{T_\gamma}$  and for  $p_{T_{\rm jet}} >  p_{T_\gamma}$
$R_{AA}^{\gamma-{\rm jet}}< 1$. In the presented transverse momentum range
the suppression at LHC is slightly larger even though the jet spectra are
harder, reflective of higher temperature, density and stopping power of the QGP
at LHC. Bands represent a range of couplings between the jet and the medium
$g_{\rm med} = 1.8 - 2.2 $. For  $p_{T_{\rm jet}} < p_{T_\gamma}$ there can be a strong enhancement
of the cross section. In the studied transverse momentum range this is the case at RHIC
and it reflects the narrower baseline jet distribution away from
 $p_{T_{\rm jet}} \approx  p_{T_\gamma}$ due to the smaller $\sqrt{s}$.
Fig.~\ref{fig:raa} illustrates the flexibility that photon-tagged afford in
comparing the properties of deconfined  strongly-interacting matter at RHIC and LHC.

Let us define by $z_{J\gamma} = {p_{T_{\rm jet}}}/{p_{T_\gamma}} $
the momentum imbalance of photon+jet events. Its distribution
can be obtained from the ${\cal O}(\alpha_{\rm em}\alpha_s^2)$ double differential
cross section ${d\sigma}/{dp_{T_\gamma}dp_{T_{\rm jet}}}$. By changing variables
from $(p_{T\gamma},p_{T_{\rm jet}})$ to
$(z_{J\gamma},p_{T_{\rm jet}})$  and integrating over $p_{T_{\rm jet}}$,  we can express:
 \begin{eqnarray}
\frac{d\sigma}{dz_{J\gamma}} = \int_{p_{T_{\rm jet}}^{min}}^{p_{T_{\rm jet}}^{max}}d p_{T_{\rm jet}}
\,  \frac{p_{T_{\rm jet}}}{z_{J\gamma}^2}
\frac{d\sigma[z_{J\gamma},p_{T_\gamma}(z_{J\gamma},p_{T_{\rm jet}})]}{dp_{T_\gamma}dp_{T_{\rm jet}}} \;.
\end{eqnarray}
Here, $p_{T_{\rm jet}}^{min}$, $p_{T_{\rm jet}}^{max}$ can be specified by the experiment and
together with the transverse momentum cuts on the isolated photons  contribute
to the shape of the $z_{J\gamma}$ distribution. In our p+p and Pb+Pb calculations at LHC
we use the CMS experimental cuts $p_{T_{\rm jet}} > 30$~GeV, $p_{T_\gamma} > 60$~GeV,
$\vert y^{\gamma} \vert < 1.44$, $\vert y^{\text{jet}} \vert < 1.6$,
$\vert \phi^{\text{jet}}-\phi^\gamma \vert > \frac{7}{8}\pi$. We implement a $k_T$ algorithm with a
radius parameter $R = 0.3$ for the jet and isolation criterion that requires
the total energy within a cone of radius $R_{\rm iso.}= 0.4$ surrounding the photon
direction to be less than 5~GeV.
In p+p and Au+Au collisions at RHIC, we consider the same cuts except $p_{T_{\rm jet}} > 10$~GeV,
$p_{T_\gamma} > 30$~GeV, and impose a different isolation cut that
requires the hadronic transverse energy in the cone of radius $R_{\rm iso.}=0.4$
around the photon direction  to be less than 7\%  of the $\gamma$ energy.

\begin{figure}[!t]
\begin{center}
\vspace*{-0.3in}
\includegraphics[width=3.5in,height=2.4in,angle=0]{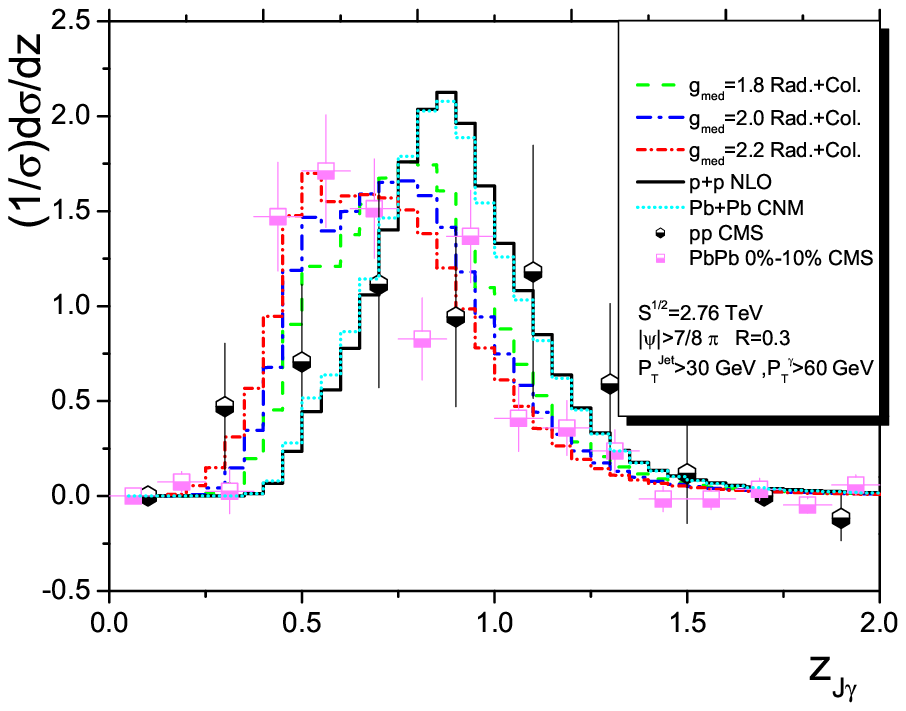} \\
\vspace*{-.3in}
\includegraphics[width=3.5in,height=2.4in,angle=0]{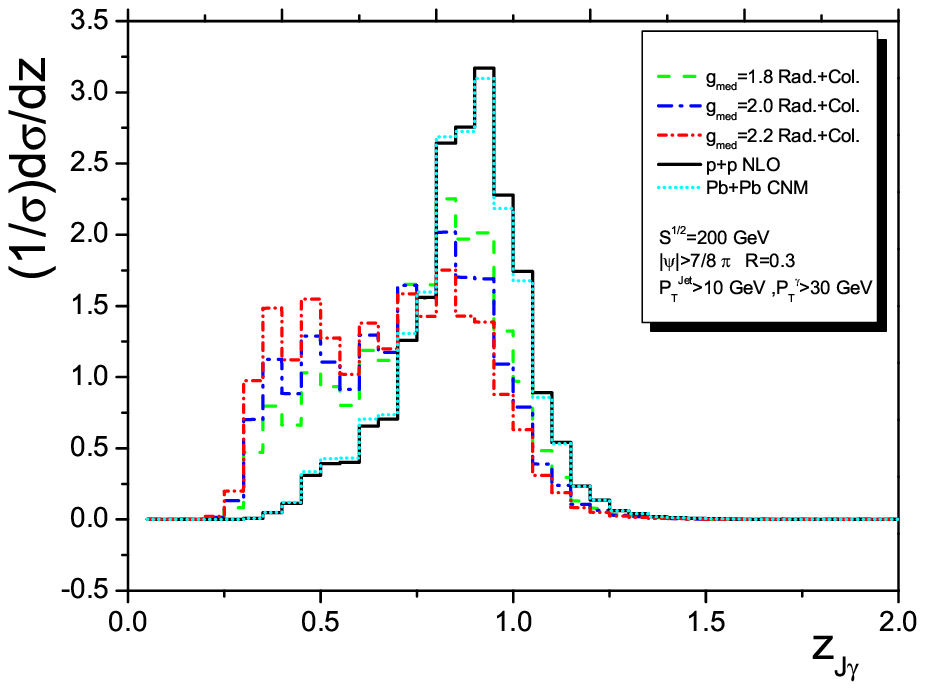}
\vspace*{-.3in}
\caption{ The isolated photon-tagged jet asymmetry distribution for different
coupling  strengths between the jet and the medium. Top panel: Pb+Pb
collisions at $\sqrt{s_{NN}}  = 2.76$~TeV. Black and magenta points are CMS
experimental data in p+p and Pb+Pb collision,
respectively. Bottom panel: Au+Au collisions at $\sqrt{s_{NN}} = 200$~GeV. }
\label{fig:illust3}
\end{center}
\end{figure}

The normalized momentum imbalance distribution $({1}/{\sigma})
{d\sigma}/{dz_{J\gamma}}$ is given in Fig.~\ref{fig:illust3}. The solid black
line shows the p+p calculation and the circles represent the CMS result with
large error bars. The dotted cyan line includes  cold  nuclear matter effects
in central Pb+Pb (top panel) and central Au+Au (bottom panel) reactions at LHC
and RHIC, respectively. These CNM effects do not affect the $z_{J\gamma}$
distribution appreciably. The physics responsible for the difference between
p+p and A+A reactions is then contained in the final-state QGP-induced parton
splitting and the dissipation of the parton shower energy in the medium. The
parameter that controls the strength of the coupling between the jet constituents
and the strongly interacting matter is $g_{\rm med}$. We investigate a
range  of values  $ g_{\rm med} = 1.8 \ ({\rm green \ dashed} )$,  $2.0 \
({\rm blue \  dot-dashed} )$, $2.2 \ ({\rm red \ short  \ dot-dashed} ) $
that has worked well in describing  the di-jet asymmetry distribution and
in predicting the inclusive jet suppression at LHC~\cite{He:2011d}.
The same range of coupling strengths has been used to predict the asymmetry
distribution of $Z^0$+jet events in heavy-ion collisions~\cite{Neufeld:2012df}.

The effect of final-state interactions is to broaden the momentum imbalance
distribution  and shift it down to smaller values of $ z_{J\gamma}$.
Our calculations include both radiative and collisional energy losses that
alter the associated jet transverse momentum.
They differ in the amount of shift observed, but the
broadening is approximately independent of the value of  $g_{\rm med}$.
It should be noted that even in p+p collisions the distribution peaks below
$z_{J\gamma}=1$.  We define the mean value of $z_{J\gamma}$ as:
 \begin{eqnarray}
     \langle z_{J\gamma} \rangle = \int dz_{J\gamma} z_{J\gamma}
\frac{1}{\sigma}\frac{d\sigma}{dz_{J\gamma}} \; ,
\end{eqnarray}
and show its values in Table~\ref{tab:table1}. The steeper falling cross
sections at RHIC energies lead not only to a narrower $z_{J\gamma}$ distribution
in  p+p collisions but also to larger broadening end shift in
$\langle z_{J\gamma} \rangle$ in A+A collisions in spite of the fact that, on
average, less energy per jet is dissipated as the parton shower forms and propagates
in the QGP. Our results, quoted in Table~\ref{tab:table1}, can also be
compared directly  to  the most central Pb+Pb data, where CMS measured the
ratio  $\langle z_{J\gamma}\rangle =
0.73\pm 0.02({\rm stat.}) \pm 0.04({\rm syst.})$~\cite{Chatrchyan:2012gt}.

\begin{table}[!t]
\caption{\label{tab:table1} Theoretical results for  $\langle z_{J\gamma}\rangle $ in
p+p, central Pb+Pb and central Au+Au reactions. Center-of-mass
energies are: LHC $\sqrt{s_{NN}}=2.76$~TeV,
RHIC $\sqrt{s_{NN}}=200$~GeV.  }
\begin{tabular}{lcr}
System& $\langle z_{J\gamma} \rangle_{\text {LHC}} $  &$\,\,
\,\,\langle z_{J\gamma} \rangle_{\text{RHIC} }$\\
\hline
p+p & 0.94  &0.90\\
A+A, CNM & 0.94 &0.89 \\
A+A, $g_{med}=1.8$ ,Rad.+Col. & 0.84 & 0.78 \\
A+A, $g_{med}=2.0$ ,Rad.+Col. & 0.80 & 0.74\\
A+A, $g_{med}=2.2$ ,Rad.+Col. & 0.71 & 0.70\\
\vspace*{-.1in}
\end{tabular}
\end{table}

In summary, we presented first results for the differential cross sections
and momentum imbalance of isolated photon-tagged jets in p+p and A+A collisions at RHIC and
LHC. We found that a theoretical approach that combines the ${\cal O}(\alpha_{\rm em} \alpha_s^2)$
perturbative cross sections with the medium-induced parton splitting  and parton shower
energy dissipation in the QGP describes quantitatively the increase of the transverse
momentum imbalance observed by the CMS experiment in central Pb+Pb reactions at
$\sqrt{s_{NN}} = 2.76$~TeV.
Through comparison between theoretical predictions, such as the modification of
the $\gamma$+jet cross sections and the associated $z_{J\gamma}$  distribution
presented here, and upcoming  experimental results, the emerging picture of in-medium
parton  shower formation  and evolution can further be tested at RHIC and  LHC.


\end{document}